\documentclass[reprint,aip,amsmath,amssymb,nofootinbib]{revtex4-2}
\usepackage{amssymb}
\usepackage{lipsum}
\usepackage{amssymb}
\usepackage{amssymb}
\usepackage{graphicx}
\usepackage{graphics}
\usepackage{amsmath}
\usepackage{placeins}
\usepackage{hyperref}
\usepackage[dvipsnames]{xcolor}
\usepackage{wasysym}
\usepackage{soul}
\usepackage{float}
\usepackage{array}
\usepackage{tabulary}
\usepackage{caption}
\usepackage{fancyhdr}
\usepackage{wrapfig}

\usepackage[normalem]{ulem}
\usepackage{graphicx}
\usepackage{color,soul}
\usepackage{subfigure}
\usepackage{dcolumn}
\usepackage{bm}
\usepackage{hyperref}
\usepackage{url}
\usepackage{multirow}
\hypersetup{
    colorlinks=true,
    linkcolor=blue,
    filecolor=blue,      
    urlcolor=blue,
    citecolor=blue,
}

\usepackage[makeroom]{cancel}

\begin{document}

\title{Evaluating Effects of Geometry and Material Composition on Production of Transversely Shaped Beams from Diamond Field Emission Array Cathodes}

    \author{\firstname{Mitchell E.} \surname{Schneider}}
	\email{schne525@msu.edu}
    \affiliation{Department of Electrical and Computer Engineering, Michigan State University, MI 48824, USA}
    \affiliation{Department of Physics and Astronomy, Michigan State University, East Lansing, MI 48824, USA}
    \affiliation{Accelerator Operations and Technology Division, Los Alamos National Laboratory, Los Alamos, NM 87545, USA}
    \author{\firstname{Heather} \surname{Andrews}}
    \affiliation{Accelerator Operations and Technology Division, Los Alamos National Laboratory, Los Alamos, NM 87545, USA}
    \author{\firstname{Sergey V.} \surname{Baryshev}}
	\email{serbar@msu.edu}
	\affiliation{Department of Electrical and Computer Engineering, Michigan State University, MI 48824, USA}
    \author{\firstname{Emily} \surname{Jevarjian}}
    \affiliation{Department of Physics and Astronomy, Michigan State University, East Lansing, MI 48824, USA}
    \author{\firstname{Dongsung} \surname{Kim}}
    \affiliation{Accelerator Operations and Technology Division, Los Alamos National Laboratory, Los Alamos, NM 87545, USA}
    \author{\firstname{Kimberley} \surname{Nichols}}
    \affiliation{Accelerator Operations and Technology Division, Los Alamos National Laboratory, Los Alamos, NM 87545, USA}
    \author{\firstname{Taha Y.} \surname{Posos}}
    \affiliation{Department of Electrical and Computer Engineering, Michigan State University, MI 48824, USA}
    \author{\firstname{Michael} \surname{Pettes}}
    \affiliation{Center for Integrated Nanotechnologies, Los Alamos National Laboratory, Los Alamos, NM 87545, USA}
    \author{\firstname{John} \surname{Power}}
    \affiliation{High Energy Physics Division, Argonne National Laboratory, Lemont, IL 60439, USA}
    \author{\firstname{Jiahang} \surname{Shao}}
    \affiliation{High Energy Physics Division, Argonne National Laboratory, Lemont, IL 60439, USA}
    \author{\firstname{Evgenya I.} \surname{Simakov}}
     \email{smirnova@lanl.gov}
    \affiliation{Accelerator Operations and Technology Division, Los Alamos National Laboratory, Los Alamos, NM 87545, USA}

\begin{abstract}
	Field emission cathodes (FECs) are attractive for the next generation of injectors due to their ability to provide high current density bright beams with low intrinsic emittance. One application of FECs worthy of special attention is to provide transversely shaped electron beams for emittance exchange that translates a transverse electron beam pattern into a longitudinal pattern. FECs can be fabricated in a desired pattern and produce transversely shaped beams without the need for complex masking or laser schemes. However, reliable and consistent production of transversely shaped beams is affected by material properties of the FEC. This paper reports the results of testing two diamond field emitter array (DFEA) FECs with the same lithography pattern and emitter geometry but different material and tip characteristics. Although both cathodes were able to sustain gradients of 44 MV/m and produce maximum output integral charge of 0.5 nC per radiofrequency (rf) pulse, their emission patterns were quite different. One cathode did not produce a patterned beam while the other one did. Differences in field emission characteristics and patterned beam production were explained by the differences in the tip geometry and the cathode material properties. The main practical takeaway was found to be that the tip sharpness was not a prerequisite for good patterned beam production. Instead, other material characteristics, such as the ballast resistance, determined cathode performance.
\end{abstract}

\maketitle

To engineer a field emission cathode (FEC) best suited for a specific application, one must consider both geometrical design and properties of the precursor material. The geometry remains the dominant ideology when useful designs are considered: the field enhancement factor which determines the FEC performance is viewed to be of a purely geometric nature. This perspective has two issues. First, field enhancement factor may change during operation of the FEC. \cite{1,2} Second, there exist efficient field emitters that have nanoscale or atomically smooth surface that therefore do not have geometrical surface field enhancement; \cite{3,4} however, they still produce electrons in the regime of field emission. To the same point, most advanced field emission materials are often non-metal and therefore the applied field does not get screened on the surface of the material due to the lack/insufficiency of free charges. Thus, studying material properties of FECs (well beyond simply searching for the lowest work function material) becomes critically important as opposed to simply utilizing the Fowler-Nordheim (FN) equation to describe any FEC as an ideal conductor with only two properties -- the field enhancement factor and the work function -- and not taking into account realistic differences between metals: semiconductors (with doping) and dielectrics.

Recently, it was demonstrated that diamond field emission array (DFEA) FECs can produce transversely shaped electron beams. \cite{5,6} In particular, production of a triangular shaped beam was demonstrated. \cite{6} However, the mechanism behind why some DFEAs can produce a shaped beam while other geometrically similar DFEAs cannot is not clear. DFEAs are periodic arrays of micron-scale nitrogen-doped micro-diamond pyramids with sharp nanometer-scale tips. In this paper, we report the results of an experiment aimed at producing shaped beams from two DFEA FECs that were tested in the same radiofrequency (rf) injector operated at L-band (1.3 GHz). An effort was made to study the effects of geometry and materials on production of shaped electron beams and directly compare the beams produced by the two cathodes that had the same geometry but different material compositions.

The two DFEA FECs used in this study are further referred to as cathode A and cathode B. Both were fabricated under the same growth conditions at the same time. Both cathodes were made in the form of a triangular pattern: multiple pyramids spaced 50 $\mu$m from each other  to form an equilateral triangle with a side of 1025 ± 2 $\mu$m. Each diamond pyramid had a base of 25 ± 1 $\mu$m and height-to-base ratio of 0.7 to 1. 

\begin{figure}[b]
	\includegraphics[width=6cm]{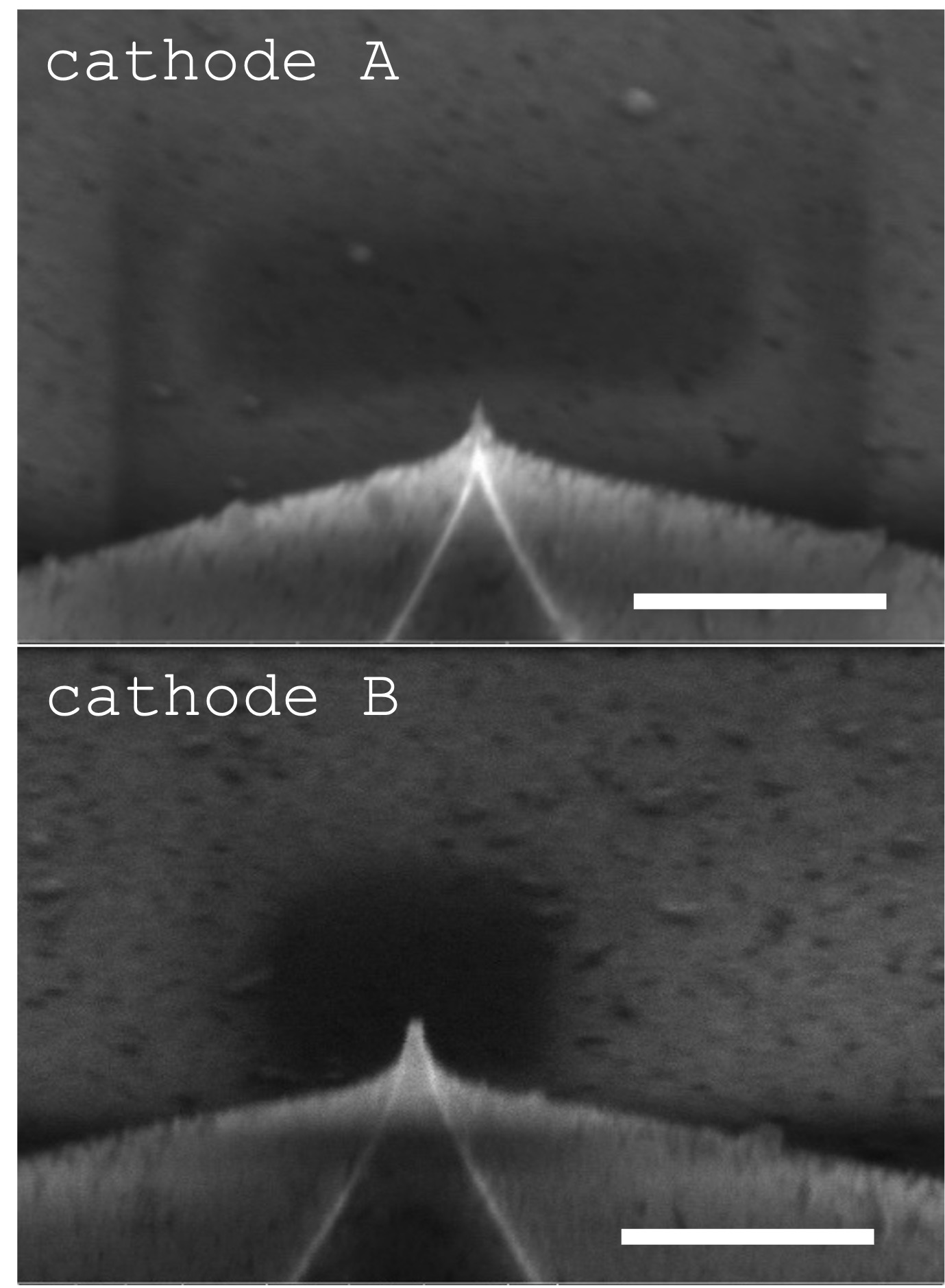}
	\caption{SEM images of two single emitter pyramids showing the sharpness of a typical emitter tip for cathode A and cathode B. The scale bar is 1 $\mu$m.}\label{fig1}
\end{figure}

The process to fabricate a DFEA cathode is explained in detail in Ref.\onlinecite{7}. Diamond is deposited into molds of inverse pyramid arrays lithographically etched in a silicon substrate. Diamond deposition was performed commercially at Advanced Diamond Technologies in Romeoville, IL. After the diamond growth was complete, the microlayer of diamond with pyramids was brazed onto a molybdenum substrate that became the actual cathode plug for the three-part assembly described in Ref.\onlinecite{8}. Small variations in the mold, such as sharpness of the edges and angles of the inverse pyramids, as well as the growth process for the diamond, led to variations in both the material properties and the geometry of the two cathodes. Scanning electron microscopy (SEM) images of the cathodes are shown in Fig.~\ref{fig1} and reveal that cathode A had much sharper, smaller diameter tips than cathode B.

The cathodes were tested and imaged in a high gradient environment in the L-band Argonne Cathode Test-stand (ACT) with an rf pulse length of 6 $\mu$s at 2 Hz repetition rate and at vacuum of approximately 10$^{-9}$ Torr. Detailed description of the ACT can be found in Ref.\onlinecite{8}.

\begin{figure}
	\includegraphics[width=8cm]{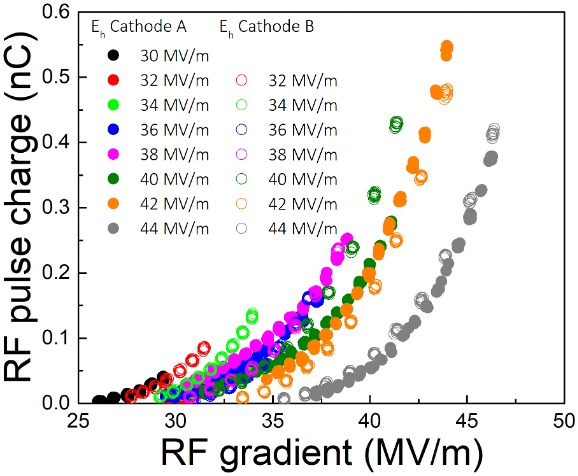}
	\caption{$Q-E$ curves for cathode A (solid circles) and cathode B (hollow circles). $E_h$ labels the highest achieved gradient for a given conditioning cycle.}\label{fig2}
\end{figure}

\begin{figure}[]
	\includegraphics[width=8cm]{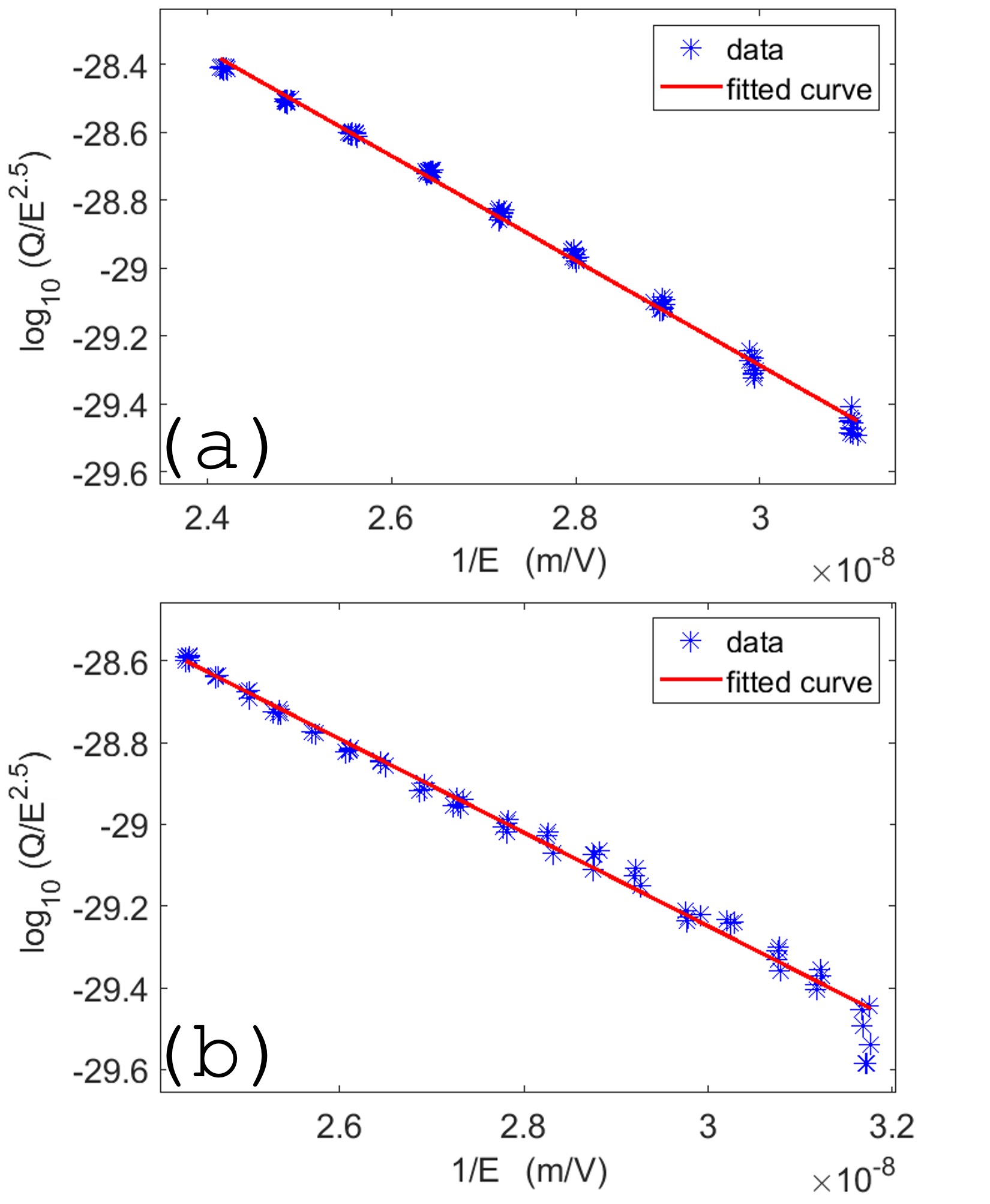}
	\caption{FN plots for (a) cathode A and (b) cathode B both conditioned to $E_h$=40 MV/m. The dependencies are linear with $R^2 > 0.99$ for each linear regression.}\label{fig3}
\end{figure}

\begin{figure*}[]
	\includegraphics[width=12cm]{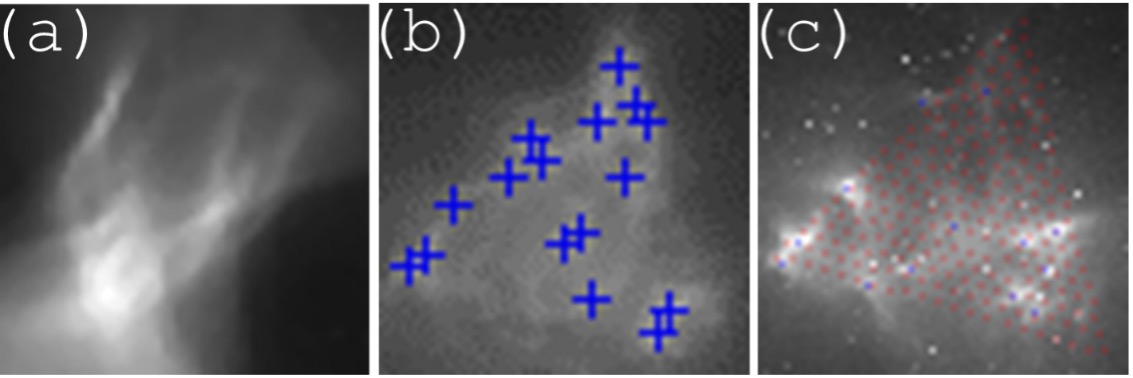}
	\caption{Image taken on YAG3 at 40 MV/m for (a) cathode A and (b) cathode B. Cathode A produced no patterned beam and a large electron cloud. Cathode B produced a triangular pattern of emission with local maxima determined using FEpic\cite{12} denoted by blue crosses (not to scale); (c) Overlay of the initial DFEA pattern times a magnification factor of 5.3 (red dots) and the local maximums determined from FEpic (blue dots).}\label{fig4}
\end{figure*}

Based on the standardized conditioning procedure presented in more detail in Ref.\onlinecite{8}, the turn-on fields were 22 MV/m for cathode A and 24 MV/m for cathode B. These turn-on fields were about 10 MV/m larger than the turn-on field measured for plane diamond cathodes in previous experiments. \cite{2,8,9} All conversions of raw data to the resulting $Q-E$ (rf pulse charge $Q$ as a function of the cathode surface gradient $E$) curves were obtained using a custom software called FEbeam.\cite{10} The $Q-E$ curves were measured for 30 MV/m and 32 MV/m for cathode A and B, respectively. At lower gradients the charge collected on the Faraday cup was below the detection threshold of the Faraday cup’s circuitry. The resulting $Q-E$ curves are shown in Fig.~\ref{fig2}. Both cathodes performed similarly, producing a maximum charge of $\sim$0.5 nC per rf pulse and achieved a maximum field of 45 MV/m before breakdown. Both showed nearly identical charge-field functional behavior and thus field enhancement factors were similar in spite of one cathode having sharper tips. The performance of both cathodes started to decay at gradients above 42 MV/m as indicated by a decrease of emitted charge. This is consistent with previous results of testing DFEAs.\cite{5} Fig.~\ref{fig3} shows the same $Q-E$ data as in Fig.~\ref{fig2} after conditioning to the gradient of 40 MV/m plotted in the FN coordinates, $log_{10}$(Q/E$^{2.5}$) vs. $1/E$, where $Q$ is the charge per rf pulse and $E$ is the gradient. The dependencies shown in Fig.~\ref{fig3} are linear, indicating that there were very little effects from space charge\cite{2} and current saturation.\cite{11}

Imaging of the electron beams produced by each cathode showed that the two cathodes performed substantially different in terms of producing a shaped beam. No shaped beam formation and only a large electron halo-like cloud was observed during the tests of cathode A (see Fig.~\ref{fig4}(a)). On the other hand, while testing cathode B, a triangle emission pattern was clearly observed (Fig.~\ref{fig4}(b)). FN analysis of emission from both cathodes indicated that the space charge did not play any significant role in beam formation from either cathode A or cathode B; therefore, the observed differences in performance of two cathodes merit further discussion.

The beam images obtained during the tests of cathode B were processed with a custom algorithm implemented in the software package FEpic\cite{12} that allows for identifying locations of emission centers. It was found that the spacing between local maxima in emission correlated well with distances between pyramids in the field emission array times a magnification factor. It was possible to overlay the observed emission pattern with the DFEA pattern photographed with an SEM. It could be seen that, if a magnification factor of 5.3 was used, the emission pattern and the SEM image overlapped fairly well (Fig.~\ref{fig4}(c)). Similar beam magnification factors were measured in previous experiments.\cite{13} This is explained by the fact that the beam emittance in the ACT rf injector is dominated by the rf field in the injector’s rf structure regardless of the cathode under test. The spacing between observed emission dots was approximately 3.25 pixels wide on the YAG screen which corresponded to the distance of 265 $\mu$m. Due to small magnification and limited camera resolution, each identified emitter appeared as a single pixel on the image. The size of the pixel of the camera is approximately 81.4 $\mu$m. If we divide that size by the magnification of 5.3 and the glow sigma factor of 2.7183 (because the glow of the YAG screen has Gaussian distribution), we obtain the upper limit on the effective size of an emitter at approximately 5.7 $\mu$m. Two other estimates of the emitter size were performed as follows.

Calculations of the capture ratio,\cite{8} done in General Particle Tracking (GPT) environment aided by custom developed field emission particle generator FEgen,\cite{14} demonstrated that the charge collection out of the gun was always close to 100\% throughout the conditioning process (i.e. at any applied gradient) as long as the emitter's radius remained larger than 100 nm. Therefore, during the experiment, we observed that charge collection must have been close to 100\% with the measured $Q-E$ curves being nearly ideally linear in F-N coordinates. It is well established that emission becomes space charge limited at current densities above 10$^7$ A/cm$^2$.\cite{15} With $\sim$1 nC charge emitted over 3 $\mu$s long flat top rf pulse,\cite{8} bunch charge emitted per each rf cycle at 1.3 GHz is close to 0.13 pC. Emission window within a single rf pulse can be between 38 and 380 ps, therefore the peak current per rf cycle $I_p$ was in the range of 0.001 to 0.01 A. Consequently, the emission area at which space charge starts playing role $A_e=I_p/10^7$ A/cm$^2$  is in the range of $10^{-10}$ to $10^{-9}$ cm$^2$. Given that space charge effects were not observed, we obtain the lower limit on the characteristic emission radius  that is 60-200 nm. Together with previous tip imaging and FN fitting results, these new estimates yield some understanding of what emitting radius and area in a DFEA pyramid are. However, most importantly geometrical analysis shed no light onto why otherwise identically fabricated DFEAs showed different emission uniformity across the array. Thus, material properties  were further analyzed to understand the difference in cathode performance.

\begin{figure*}
	\includegraphics[height=5cm]{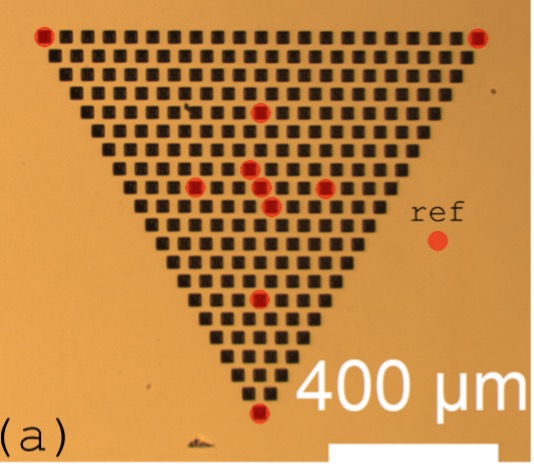}\hspace{2cm}\includegraphics[height=5cm]{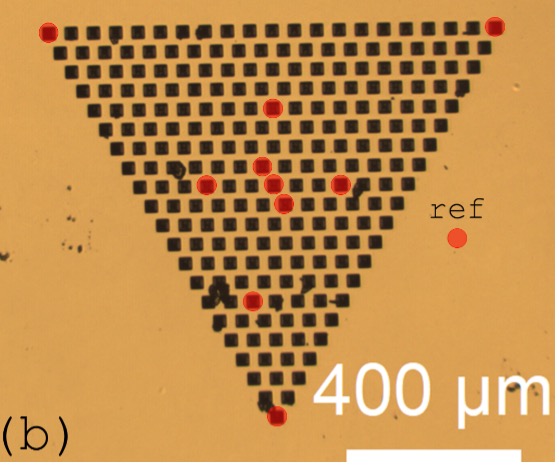}
	
	\includegraphics[width=8cm]{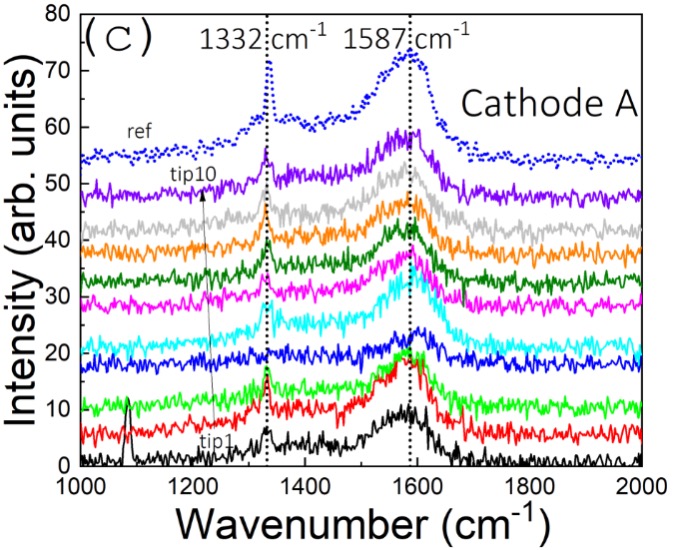}\hspace{.5cm}\includegraphics[width=8cm]{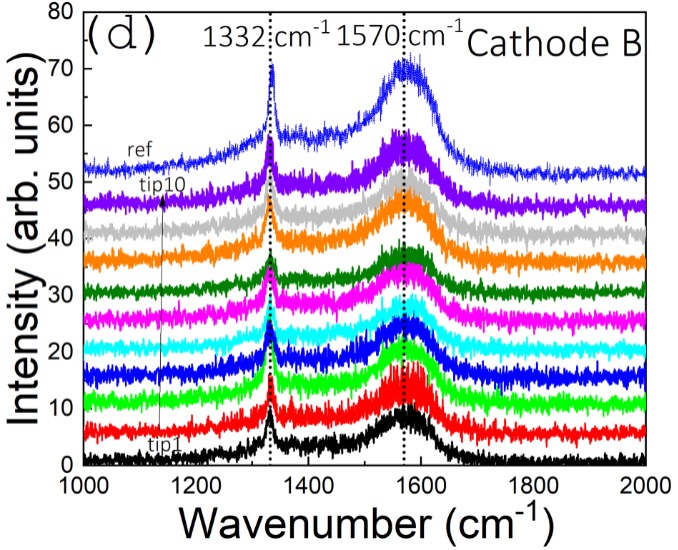}
	
	\caption{Optical images of (a) cathode A and (b) cathode B with tips that underwent Raman spectroscopic studies after fabrication labeled with red circles. Corresponding Raman spectra for (c) cathode A and (d) cathode B are presented with respect to the base reference signal labeled as \textbf{ref}.}\label{fig5}
\end{figure*}

Raman spectroscopic mapping was employed to gain insight into material properties of both cathode A and cathode B that may explain different performances of two cathodes. Deep ultraviolet (DUV) confocal Raman spectroscopy was performed with sub-micron lateral resolution using a 244 nm probing laser light source. Raman spectra were taken on 10 selected pyramids and the base of the diamond substrate (reference signal) to evaluate graphitic-to-diamond content in both samples: any polycrystalline diamond is a mixed phase material and Raman spectroscopy resolves one phase from another.\cite{16} Figs.~\ref{fig5} (a) and (b) show which pyramids were evaluated in each array. The Raman spectra (Fig.~\ref{fig5}(c) and (d)) confirmed that both DFEAs consisted of pyramids made out of nanodiamond\cite{7} that show on the Raman spectra as two prominent peaks. The peak related to the diamond $sp^3$ phase is centered at 1332 cm$^{-1}$ and is called the D peak. Unlike in single crystalline diamond, there is also a large, broad peak peak in the spectra of both cathodes slightly below the wavelength of 1600 cm$^{-1}$. This is called the G peak and reflects the presence of carbon grain boundary $sp^2$ phase. Sampling different pyramids across the array determined that cathode B had much better uniformity of the intensities and positions of both D and G peaks across the array (Fig.~\ref{fig5}(d)). On the other hand, cathode A had a large variation in the diamond $sp^3/sp^2$ composition across the array, which can be seen from Fig.~\ref{fig5}(c) with varying intensities and spectral positions of both D and G peaks most pronounced for the blue curve in Fig.~\ref{fig5}(c) (marked with an arrow) for which the diamond D peak completely disappeared and the G peak upshifted to 1600 cm$^{-1}$ indicating that the particular location of measurement mostly consisted of phase segregated nanographite. Overall, cathode A was stronger graphitized: all Raman measurements across cathode A had the location of G peak between 1585 and 1600 cm$^{-1}$, while G peak for cathode B was found consistently at 1570 cm$^{-1}$ with no observable variation. Lower wavenumbers for the positions of G peak for cathode B also suggested that the grain boundary phase was more amorphous for this cathode. Based on our previous controlled experiments,\cite{17} it can be concluded that cathode B should have had a higher resistivity than cathode A. Furthermore, we can conclude that the tip material quality of the respective cathode is nearly identical to that of its base. Thus, perhaps, the unexpected/uncontrollable variations in nanodiamond deposition methodology (rather than post-processing fabrication steps) are responsible for making cathodes A and B different.

\begin{figure}
	\includegraphics[width=8cm]{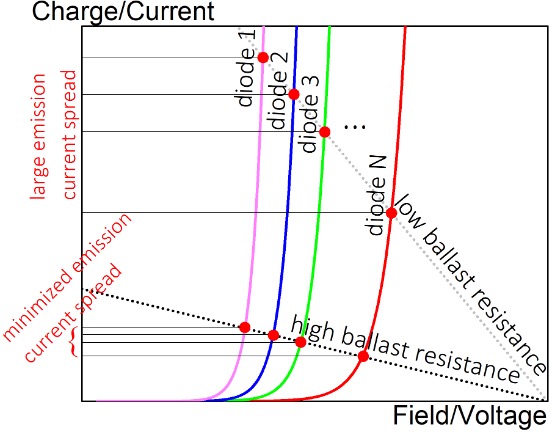}
	\caption{Simplified load-line representation of the ballast resistance effect.}\label{fig6}
\end{figure}

It can be seen from the collected Raman spectra that the sample that successfully produced a triangularly shaped beam (cathode B) most likely had a better emission uniformity across the array that was driven by a better material quality and uniformity. Possibly even more important, cathode B had larger resistivity thus translating into a larger resistance. Therefore, we propose that the so-called built-in ballast resistance effect\cite{18,19} helped stabilize operation of cathode B and allowed for more uniform local diode emission characteristics. Fig.~\ref{fig6} illustrates this effect. Several diode exponential curves represent emission curves of several tips in the DFEA array. In the load line representation, the ballast resistance can be seen as a negative slope line, where the line angle is the reverse of the resistance. For the low ballast resistance (such as was likely the case in cathode A), the ballast resistance line crosses the diode characteristics at larger emitted currents with a larger spread in currents produced by each pyramid. For the higher ballast resistance (such as was likely the case in cathode B), the ballast resistance line crosses the non-linear diode characteristics at lower emitted currents with a much smaller spread of currents produced by different pyramids. Since the total charge produced by both cathodes was approximately the same at 0.4-0.5 nC (Fig.~\ref{fig2}), this means that more pyramids were emitting in cathode B, as compared to cathode A, thus yielding better emission uniformity and hence the shaped beam. In extreme case with very low ballast resistance, at low voltage/field only one pyramid would turn on and emit all the current. As the field increased, other pyramids would start turning on, but the current from the pyramid that started emitting first would be extraordinarily high. This emitter is likely to burn fast due to thermal damage. On the contrary, in a cathode with a higher ballast resistance smaller current would be produced by each single pyramid, but all the tips would emit at the same time providing the same amount of total current/charge. If geometry (i.e. tip sharpness) plays role, this role could be secondary. This result is consistent with previous results by Jarvis $et$ $al.$ \cite{20} who showed that DFEAs performed best after conditioning process that would turn initially sharp tips duller.

\begin{figure*}
	\includegraphics[height=5cm]{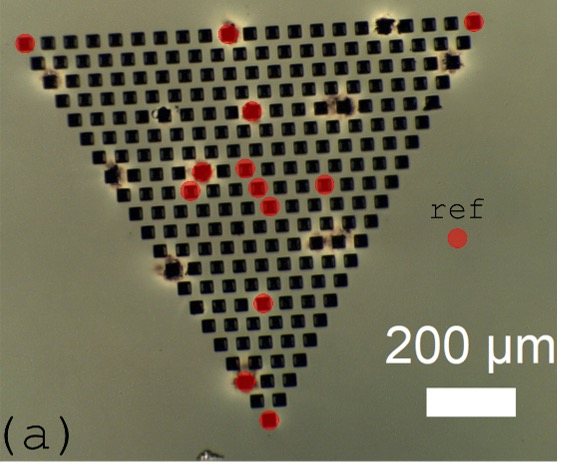}\hspace{2cm}\includegraphics[height=5cm]{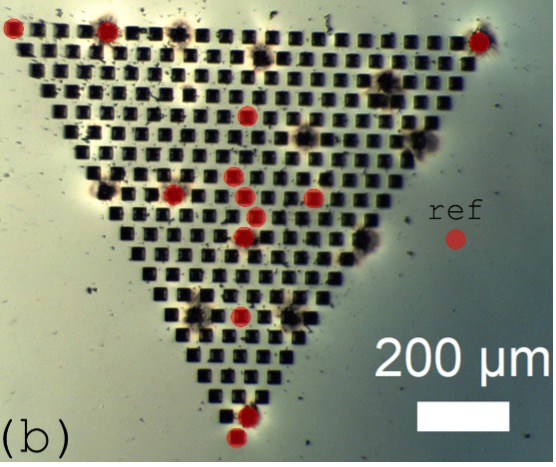}
	
	\includegraphics[width=8cm]{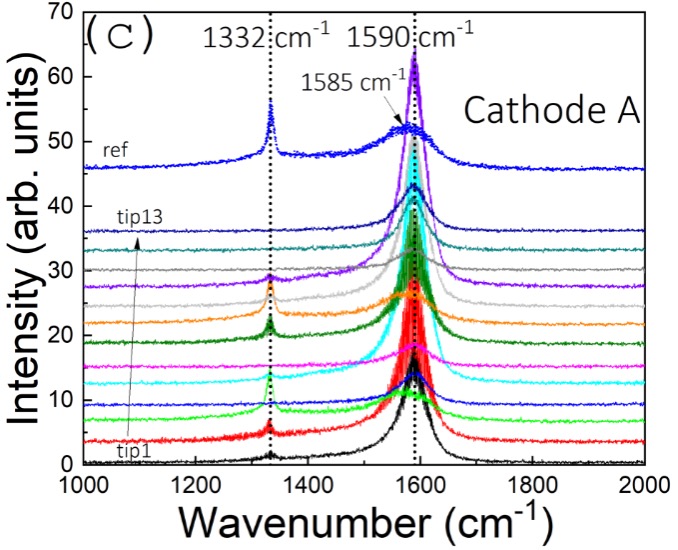}\hspace{.5cm}\includegraphics[width=8cm]{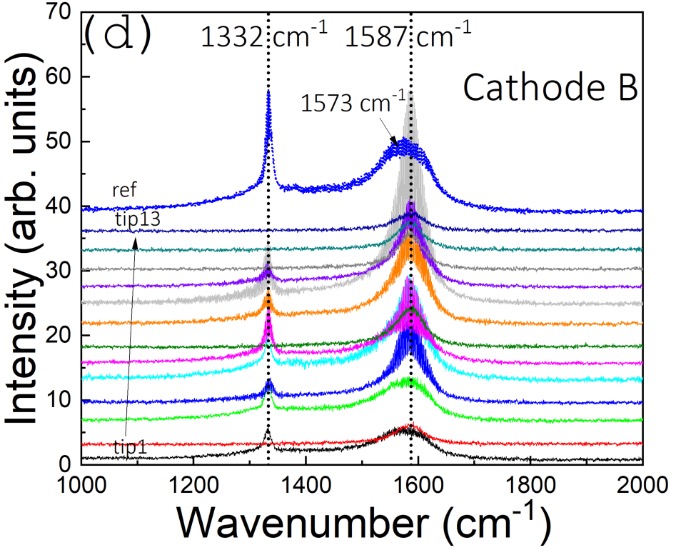}
	
	\caption{Optical images of (a) cathode A and (b) cathode B with tips that underwent Raman spectroscopic studies after high power rf testing are labeled with red circles. Corresponding Raman spectra for (c) cathode A and (d) cathode B are presented with respect to the base reference signal labeled as \textbf{ref}.}\label{fig7}
\end{figure*}

Lastly, post-mortem Raman analyses (Fig.~\ref{fig7} showing localized breakdowns at the array tip locations) enabled another set of important and consistent conclusions. One can see that the base reference signal before and after high power tests remained unchanged (compare curves labeled with ``ref'' in Fig.~\ref{fig5}(c) and 5(d) versus Fig.~\ref{fig7}(c) and ~\ref{fig7}(d)) both in terms of D peak and G peak positions and relative amplitudes. However, signals taken from the tips revealed significant material modification. For almost every line, the D peak at 1332 cm$^{-1}$ had significantly lower intensity (or was even absent) suggesting decrease in $sp^3$ content of the material in the tips and its conversion into $sp^2$ (graphitization phase transition) caused by the heat generated by emission current. This process is synchronous with the G peak (related crystalline graphite) shifting to higher wavenumbers close to 1600 cm$^{-1}$ and getting narrower. This effect is especially pronounced for the cathode B for which the G peak moved from 1570 to 1590 cm$^{-1}$. In the tips, diamond $sp^3$ content converted to graphitic $sp^2$ content. Altogether, $sp^2$ content crystallized stronger as the cathode operated likely due to heat generation.\cite{8,21} Both Fig.~\ref{fig7}(c) and ~\ref{fig7}(d) have some curves with no visible D peak. These curves correspond to the tips that exploded: the entire tip got converted into nanocrystalline graphite. This is fully consistent with our previous findings of self-driven glow discharge and arc formation in a nanodiamond field emission diode.\cite{21} Thus, conditioning and operation can lead to two scenarios: either smooth/stable or explosive runaway $sp^3$ to $sp^2$ conversion. The "smooth conditioning" is feasible if the current load per tip is minimized while allowing more tips to support the necessary output charge, that is via the ballast resistance engineering.

In conclusion, two cathodes, identical from the fabrication point of view, were tested in an L-band rf injector with the goal to produce a transversely shaped beam and for testing reliability and reproducibility. It was found that, although the two cathodes performed identically in terms of generating the same amount of charge at the same macroscopic injector gradient, cathode B produced the required triangularly shaped beam while cathode A did not. Comparative analysis based on electric properties and in situ imaging led to the conclusion that geometrical properties could not be the primary or, at least, sole reason for such a drastic difference in performance. DUV Raman spectroscopy revealed significant differences in intrinsic material quality: cathode A, had higher graphitic content (compared to diamond content) and higher degree of segregation than cathode B. We explained the observed differences using the concept of ballast resistance. Cathode A, with low ballast resistance, had less emission uniformity as compared to cathode B. Future work will be aimed at developing further insight into and controlled implementations of the ballast resistance effect. Potentially, this will lead to better emission uniformity in  future cathodes for high power rf applications that will be able to achieve simultaneous emission stability and uniformity. The ultimate goal is to achieve high production yield of cathodes with predicable emission properties.

\section*{Acknowledgments}
The work at Los Alamos National Laboratory (LANL) was funded by the Laboratory Directed Research and Development program under project No. 20180078ER and 20210640ECR. This work was performed, in part, at the Center for Integrated Nanotechnologies, an Office of Science User Facility operated for the U.S. Department of Energy (DOE) Office of Science. LANL is operated by Triad National Security, LLC, for the National Nuclear Security Administration of U.S. Department of Energy (Contract No. 89233218CNA000001).

The work at AWA was funded through the U.S. Department of Energy Office of Science under Contract No. DE-AC02-06CH11357.

The work by Mitchell Schneider was supported by LANL was funded by the LANL Laboratory Directed Research and Development program under project No. 20180078ER. The work by Emily Jevarjian, Taha Posos and Sergey Baryshev was supported by the U.S. Department of Energy, Office of Science, Office of High Energy Physics under Award No. DE-SC0020429.

\bibliography{DFEA}

\end{document}